\documentclass[conference]{IEEEtran}
\IEEEoverridecommandlockouts
\usepackage{cite}
\usepackage{amsmath,amssymb,amsfonts}
\usepackage{bm}   
\usepackage{algorithmic}
\usepackage{graphicx}
\usepackage{textcomp}
\usepackage{tikz}
 \usepackage{pgfplots}
 \pgfplotsset{compat=newest}
  \usetikzlibrary{plotmarks}
  \usetikzlibrary{arrows.meta}
  \usepgfplotslibrary{patchplots}
\usetikzlibrary{shapes,arrows}
\usetikzlibrary{decorations.pathmorphing}
\usetikzlibrary{matrix,positioning}
\usetikzlibrary{backgrounds}
\usepackage[]{algorithm2e}
\usepackage{algorithmic}
\usepackage{program}  
\usepackage{mathtools}

\usetikzlibrary{arrows.meta,
                calc, chains,
                decorations.pathreplacing,
                calligraphy,
                positioning}
                
\usepackage{xcolor}
\usepackage{xstring}
  
\usepackage{scalerel}
\usepackage{cuted}
\usepackage[font=small,labelfont=bf,tableposition=top]{caption}
\usepackage{blindtext}

\usepackage{booktabs}
\usepackage{siunitx}
\usepackage{tabularx, makecell}
\usepackage{multirow}

\def\BibTeX{{\rm B\kern-.05em{\sc i\kern-.025em b}\kern-.08em
    T\kern-.1667em\lower.7ex\hbox{E}\kern-.125emX}}
    
\usetikzlibrary{calc, spy}
\pgfplotsset{compat=1.13}
\usetikzlibrary{calc,patterns,angles,quotes}
  \usetikzlibrary{plotmarks}
  \usetikzlibrary{arrows.meta}
  \usepgfplotslibrary{patchplots}
\usetikzlibrary{shapes,arrows}
\usetikzlibrary{decorations.pathmorphing}
\usetikzlibrary{matrix,positioning}
\usetikzlibrary{backgrounds}
\usetikzlibrary{arrows.meta,
                calc, chains,
                decorations.pathreplacing,
                calligraphy,
                positioning}
\usepackage{subcaption}
 
\renewcommand{\vec}[1]{{\bm #1}}
\newcommand{\matr}[1]{{\mathbf #1}}
\newcommand{\msub}[1]{\scaleto{#1}{2pt}}
\newcommand{\msubl}[1]{\scaleto{#1}{4pt}}

\long\def\comment#1{}

\newfont{\bbb}{msbm10 scaled 700}

\newfont{\bb}{msbm10 scaled 1100}

\newcommand{\av}{{\bf a}}

\newcommand{\cv}{{\bf c}}

\newcommand{\fv}{{\bf f}}

\newcommand{\zv}{{\bf z}}

\newcommand{\Hm}{{\bf H}}

\newcommand{\Nc}{{\cal N}}

\renewcommand{\vec}{{\rm vec}}

\begin{document}

\title{Adaptive Beam Tracking based on Recurrent Neural Networks for mmWave Channels\\
\thanks{The Work of S.K. Dehkordi was funded by Bundesministerium für Bildung und Forschung (BMBF) within the project ForMikro-6GKom. }
}

\author{\IEEEauthorblockN{Saeid K.Dehkordi}
\IEEEauthorblockA{\textit{Dept. of Information Theory} \\
\textit{Technische Universität Berlin}\\
Berlin, Germany \\
s.khalilidehkordi@tu-berlin.de}
\and
\IEEEauthorblockN{Mari Kobayashi}
\IEEEauthorblockA{\textit{Institute for Communications Engineering} \\
\textit{Technische Universität München}\\
Munich, Germany \\
mari.kobayashi@tum.de}
\and
\IEEEauthorblockN{Giuseppe Caire}
\IEEEauthorblockA{\textit{Dept. of Information Theory} \\
\textit{Technische Universität Berlin}\\
Berlin, Germany \\
caire@tu-berlin.de}

}

\maketitle

\begin{abstract}

The performance of millimeter wave (mmWave) communications critically depends on the accuracy of beamforming both at base station (BS) and user terminals (UEs) due to high isotropic path-loss and channel attenuation. 
In high mobility environments, accurate beam alignment becomes even more challenging as the angles of the BS and each UE must be tracked reliably and continuously.  
In this work, focusing on the beamforming at the BS, we propose an adaptive method based on Recurrent Neural Networks (RNN) that tracks and predicts the Angle of Departure (AoD) of a given UE. Moreover, we propose a modified frame structure to reduce beam alignment overhead and hence increase the communication rate. Our numerical experiments in a highly non-linear mobility scenario show that our proposed method is able to track the AoD accurately and achieve higher communication rate compared to more traditional methods such as the particle filter. 
\end{abstract}


\section{Introduction}

Millimeter wave (mmWave) communication systems, operating in frequency bands of 30-300 GHz, are considered as a promising technology for 5G and beyond cellular systems to achieve a high data rate thanks to wide frequency bands \cite{6G}. Due to the large isotropic path loss, large antenna arrays are typically deployed at base station (BS) and/or user terminals (UEs) in order to form narrow beams between BS-UE pairs. The classical beam sweeping mechanism is highly inefficient as its complexity increases substantially with the number of antenna arrays and its practical implementation using quantized phases may limit its accuracy. Therefore,  a number of low-complexity schemes for initial beam alignment have been proposed in the literature (see e.g. a compressed sensing approach in \cite{X.Song_OSPS} and references therein).
However, the existing schemes cannot be adapted directly to the high-mobility scenario, where the displacement of UEs may significantly increase the probability of beam misalignment. In such a scenario, the angle of each UE shall be estimated continuously through beam tracking methods (e.g. \cite{BeamTracking,Va, Love, PF} and references therein).  
For example, \cite{Va, Love, PF} proposed variants of Kalman filter (KF) to address this problem.  
Since their restricted channel model together with high complexity makes these methods \cite{Va, Love} impractical in high mobility environments, \cite{PF} considered the use of particle filtering (PF) in a time-varying channel and demonstrated an improved performance by modeling the non-linearities of the channel compared to other KF variants. Further, \cite{Palacios} attempted to overcome the computational burden.

In this work, by focusing on the beamforming at the BS, we propose a beam tracking approach based on Recurrent Neural Networks (RNN) that can be applied to any arbitrary UE mobility pattern. The proposed method takes advantage of the temporal correlations within measurement data and learns the amount of \textit{adjustment} required for the beam direction from 
the sequence of measurements by approaching the problem as a classification task. 
Based on the proposed scheme, we define a modified frame structure with variable length which can be adapted to reduce overhead. The numerical examples in a highly non uniform linear motion scenario using the Quadriga channel generator demonstrate that our proposed scheme can track the AoD accurately and achieve higher communication rate compared to the PF.  

Finally, we remark that a number of recent works extensively adapted machine/deep learning for beamforming design over the mmWave channels (see e.g. \cite{Lim, Guo,Elbir1, Sim,Huang}). In \cite{Elbir1}, a convolutional neural network (NN) is proposed for an optimal beamformer design. A deep learning framework was proposed for beam selection by using the channel state information (CSI) in \cite{Sim}, while \cite{Liu} demonstrates an auto-encoder/decoder architecture for robust angle estimation.  Although \cite{Lim, Guo} also addressed beam tracking using machine learning tools, the proposed method differs from these approaches. In \cite{Lim}, only fully connected layers, unable to capture time correlations of input data, are considered. The work of \cite{Guo} aims to infer the correct beam index by predicting (i.e. regression problem) the channel vector under the assumption of a linear motion path. More recently, the work in \cite{Qlearning} applied a Reinforcement Learning principle for UAV beam tracking, where the authors deal with the design of the reward function by introducing thresholds for the reward. These threshold values need to be optimized depending on the operational SNR. In contrast, our model does not require any SNR specific design parameter. While a comparison of the NN based methods is an interesting topic for future work, due to the limited space, we have restricted our comparison with the well assessed and understood PF scheme.

%

\section{System Model}
\subsection{Frame Structure and Signaling Scheme}\label{measurement_model}
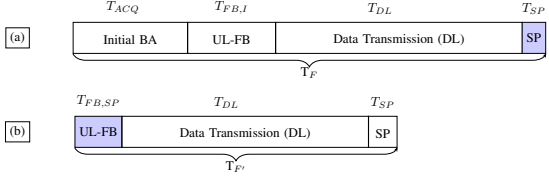
\begin{figure}[h]
\vspace{-0.3cm}
\centering
\scalebox{.5}{
%
%
%
%
%
%
%
%
%
%
%
%


\usetikzlibrary{arrows.meta,shapes.arrows,chains,decorations.pathreplacing}

\newlength\myht
\settoheight{\myht}{$n-2$}
\tikzset{%
  MyStyle/.style={draw, text width=80pt, text height=10pt, text centered,minimum height=\myht+2*3*1mm)},
  MyStylef/.style={draw, text width=60pt, text height=10pt, text centered,minimum height=\myht+2*3*1mm)},
  MyStyleD/.style={draw, text width=180pt, text height=10pt, text centered,minimum height=\myht+3*2*1mm)},
  MyStyleE/.style={draw, text width=15pt, text height=10pt, text centered,minimum height=\myht+2*3*1mm)},
  MyStyleS/.style={draw, text width=40pt, text height=10pt, text centered,minimum height=\myht+2*3*1mm)},
  boxy/.style = {rectangle,draw=black, minimum height=\myht+2*3*1mm ,fill=blue!20},
  boxy2/.style = {rectangle,draw=black, minimum height=\myht+2*3*1mm ,fill=red!20},
  myarrow/.style={shape=single arrow, rotate=90, inner sep=5pt, outer sep=0pt, single arrow head extend=0pt, minimum height=7.5pt, text width=0pt, draw=blue!50, fill=blue!25}
}

\begin{tikzpicture}[-{Stealth[length=2.5pt]}]
  \begin{scope} [shift={(-2,0)},name=scope1,start chain, node distance=-.5pt]
    \foreach \name [count=\xi] in {1}{ \node[MyStyle, on chain] (vlos\xi) {Initial BA};}
     \foreach \name [count=\xi] in {1}{ \node[MyStylef, on chain] (vlosf\xi) {UL-FB};}
     \foreach \name [count=\xi] in {1}{ \node[MyStyleD, on chain] (vlosD\xi) {Data Transmission (DL)};}
     \foreach \name [count=\xi] in {1}{ \node[boxy, on chain] (vlosE\xi) {SP};}
  \end{scope}
  \node[above= 0.1cm of vlos1, xshift=-0.2cm]{$T_{ACQ}$};
   \node[above= 0.1cm of vlosf1, xshift=-0.0cm]{$T_{FB,I}$};
    \node[above= 0.1cm of vlosD1, xshift=-0.5cm]{$T_{DL}$};
    \node[above= 0.1cm of vlosE1, xshift=0.0cm]{$T_{SP}$};

  \draw [decorate,decoration={brace,amplitude=10pt,mirror}]
  (vlos1.south west) -- (vlosE1.south east) node[black,midway,below=7pt]
  {T$_F$};
%

  \begin{scope} [shift={(-2.85,-2.5)}, start chain, node distance=-.5pt]
    \foreach \name [count=\xi] in {2}{ \node[boxy, on chain] (vlosf\xi) {UL-FB};}
     \foreach \name [count=\xi] in {2}{ \node[MyStyleD, on chain] (vlosD\xi) {Data Transmission (DL)};}
     \foreach \name [count=\xi] in {2}{ \node[MyStyleE, on chain] (vlosE\xi) {SP};}
  \end{scope}
    \node[above= 0.1cm of vlosf1, xshift=-0.0cm]{$T_{FB,SP}$};
    \node[above= 0.1cm of vlosD1, xshift=-0.5cm]{$T_{DL}$};
    \node[above= 0.1cm of vlosE1, xshift=0.0cm]{$T_{SP}$};
    
      \draw [decorate,decoration={brace,amplitude=10pt,mirror}]
  (vlosf1.south west) -- (vlosE1.south east) node[black,midway,below=7pt]
  {T$_{F'}$};
  
  \node[draw] at (-5, 0)   (a) {(a)};
  \node[draw] at (-5, -2.5)   (b) {(b)};
\end{tikzpicture}

%
}
\caption{Frame structures depending on operating mode. Note, the UL-FB slot at the beginning receives feedback from the previous SP.}
\label{fig:frame_strc}
\end{figure}

Fig.~\ref{fig:frame_strc} illustrates the proposed frame structure consisting of two different types depending on the operating mode: (a) initialization frame and (b) secondary frame. Frame (a) is used at the initialization of the system. During initial beam alignment (BA) phase, a beam alignment takes place where the BS transmits a number of pilots to sweep the beam space. The BS receives feedback (FB) signal from the UE on the previously transmitted pilots on the uplink (UL) channel, in the  \textit{UL-FB} phase, as explained in Section \ref{initial_spectrum}. During the \textit{Data Transmission} phase, the data is transmitted in downlink (DL) channel. The final phase  refers to Secondary Probing \textit{SP} phase of variable length dedicated to transmit a reduced number of pilots for further channel probing. 
Frame (b) of length $T_{F'}<T_F$ comes into operation once a reliable connection between a BS-UE pair is established after the initialization frame (a). During the first phase \textit{UL-FB}, the BS receives feedback on the previous probing (SP) from UEs. The second and third phases are identical to that of the initialization frame (a).      

\vspace{-0.3cm}
\subsection{Transmission Scheme and Measurement Equation}\label{measurement_model}
Suppose the transmitter (BS), with $N_{\msub{TX}}$ uniformly and linearly positioned elements, sends pilot symbols to the receiver receiver (UE), with $N_{\msub{RX}}$ elements over 
the time-varying MIMO channel at frame $k$ by 
\begin{align}
\Hm_k(t, \tau) = \alpha_k\av_{\msub{RX}}(\varphi_k)\av_{\msub{TX}}^{H}(\theta_k)  \delta(t-\tau_k) e^{j2\pi\nu_k t}
\end{align}
where $\alpha_k, \tau_k, \nu_k$ denotes
the attenuation coefficient, the delay, and the Doppler shift, respectively, $\{d_i\}$ is the antenna spacing with $d_1=0$, $\lambda$ is the wavelength, 
$\av_{\msub{TX}}(\theta_k)$ is a steering vector at the Tx given by 
\begin{equation}
\av_{\msub{TX}}(\theta_k)  = [ e^{j\frac{2\pi}{\lambda}d_{1}\cos(\theta_k)},... , e^{j\frac{2\pi}{\lambda}d_{N_{\msub{TX}}}\cos(\theta_k)}]^{T}\
\end{equation}
for the angle-of-departure (AoD) denoted by $\theta_k$, $\av_{\msub{RX}}(\varphi_k)$ is defined similarly for the angle-of-arrival (AoA) denoted by $\varphi_k$. 
Since we consider a LoS path between the UE/BS, we have $\varphi_k=\theta_k$. The model above can be easily extended to multiple paths. 
 The parameters $(\theta_k, \alpha_k, \tau_k, \nu_k)$ remain constant over a frame duration. 

We consider the beamforming vector $\fv$ and the combining vector $\cv$ according to a predefined beam codebook. Namely,  by focusing on the beamforming codebook at the BS side, denoted by $\mathcal{CB}=\{\fv_1, \dots, \fv_G\}$, where we let $\fv_i =\fv(\tilde{\theta}_i)= \frac{1}{\sqrt{N_{\msub{TX}}}}\av_{\msub{TX}} (\tilde{\theta}_i),  i\in \{1, \dots, G\}$ for some angles $\tilde{\theta}_i$. Then, the observed signal at the UE for frame $k$ after the combining vector is given by  
\begin{equation}\label{measurements}
z_k(t) = \alpha_k\cv^{H}(\tilde{\varphi})\av_{\msub{RX}}(\varphi_k)\av_{\msub{TX}}^{H}(\theta_k)\fv(\tilde{\theta}) x(t-\tau_k) e^{j2\pi\nu_k t} +\eta(t)
\end{equation}
where $\eta(t) \sim \Nc(0,\sigma_{\eta}^{2} )$ is the additive Gaussian noise.

 The objective of initial BA is to acquire an accurate AoA estimate $\tilde{\theta}\in \{\tilde{\theta}_1, \dots, \tilde{\theta}_G\}$ 
 so that the corresponding beamforming vector $\fv(\tilde{\theta})$
  is used for the DL data transmission.  
 We also assume that the codebooks available for UL and DL operation at the BS are identical.  

\subsection{Initial Channel State Estimation}\label{initial_spectrum}
In order to focus on tracking the UE's angular location at the BS side, we further assume that the UE is equipped with a single antenna. The UE initially listens to the channel during the initial BA period 
where the BS sends pilots through $T$ probing directions according to the predefined codebook $\mathcal{CB}$. 
By replacing $\fv(\tilde{\theta})$ with $\fv_i$ for $i=1, \dots, T$ in \eqref{measurements} followed by suitable sampling every $T_{ACQ}/T$, we obtain $T$ discrete observations or measurements, denoted by $\zv= \{z_{\tilde{\theta}_{1}}[1],...,z_{\tilde{\theta}_{T}}[T]\}$, where $z_{\tilde{\theta}_{i}}[i]$ is the received signal when probing in the direction $\tilde{\theta}_{i}$ at the $i^{th}$ sampling instant. After this initial BA phase, the UE selects the best beam direction $\tilde{\theta}_m$ corresponding to $\text{max}_{i=1,...,T}(z_{\tilde{\theta}_{i}}[i])$ and feeds back $\zv$ to the BS during UL-FB phase.   

\section{Beam Tracking Methods}

\subsection{Classical Beam Tracking}\label{Classical}
Beam tracking based on Bayesian statistical inference principles have been extensively investigated in literature\cite{Va, Love, PF}. In this subsection, we provide a brief overview of the PF technique similar to that described in \cite{PF}. 
The filtering problem consists of estimating the internal states in dynamical systems when partial observations are available in the presence of random perturbations. The objective is to compute the posterior distributions of the states of a Markov process, given some noisy and/or partial observations. Based on the system model described in (\ref{measurements}) state is sequentially estimated by using the measurement $z_{\tilde{\theta}_m}[k]$ (for brevity $z_k$) at each time frame. Due to the nonlinear dependency between the state space and the measurements in the problem at hand, the PF is particularly well suited among Kalman filtering methods. By letting  $\tilde{\theta}[k], z_k $ denote the state and the observation at frame $k$ obtained from (\ref{measurements}) after a suitable sampling\footnote{It is also possible to consider multiple measurements for a given frame.}, we model the transition of the state as $\tilde{\theta}[k] =\tilde{\theta}[k-1] + u_k$ where $u_k\sim \Nc(0, \sigma_{\tilde{\theta}}^2)$ is the process noise (i.e. perturbation). 
%
We can modify \cite{PF} by focusing on particles (beams) within a reduced region of interest of the beam space. These regions can be selected adaptively, for instance in the case of static or slow moving users, the space in proximity of the main beam direction of the previous step will have a denser distribution of particles. 

\subsection{Proposed method with Recurrent Neural Networks}
The particle filter discussed previously incurs increased complexity and overhead in the massive MIMO regime and in high mobility channels. 
In this section we consider a RNN approach for the aforementioned tracking problem by exploiting the time correlation between measurement data. Since the underlying channel parameters such as the AoD, the range, and the Doppler shift in (\ref{measurement_model}) evolve in time with some memory, 
a recurrent NN architecture seems a viable solution for this problem. In recent years, Long Short-Term Memory (LSTM) networks have been used to create large recurrent networks that in turn can be used to address complicated sequence problems in machine learning and achieve state-of-the-art results. The LSTM network, is trained using backpropagation through time and overcomes the vanishing gradient problem in this way. LSTM networks have memory blocks that replace neurons which are connected through layers. A block contains gates that manage the block’s state and output, thus developing a memory state for recent sequences. A block operates upon an input sequence and each gate within a block uses the sigmoid activation units to control whether they are triggered or not, making the change of state and addition of information flowing through the block conditional.

\begin{figure}[h]
\centering
\scalebox{.45}{\begin{tikzpicture}
    [
        box/.style={rectangle,draw=black, ultra thick, minimum size=1cm},
    ]

\tikzset{
  boxy/.style = {rectangle,
    draw=black,
     ultra thick, 
     minimum size=1cm ,
     fill=blue!30
  }
}

\newcommand{\mywind}[1]{\node [mywind] }

\foreach \x/\y in {0/0, 1/1,2/,3/ ,4/ ,5/ ,6/ ,7/ ,8/ ,9/G-1,10/G}
        \node[box] at (\x,0){\y};
        
        \foreach \x/\y in {3/m-L,4/,5/k,6/,7/m+L}
        \node[boxy] at (\x,0){\y};
        
        \foreach \x/\y in {3/0,4/0,5/0,6/1,7/0}
        \node[box] at (\x,-3){\y};

\draw[decorate,decoration={brace,mirror},thick] (2.5,-.7) -- node[below]{sliding feature window (input features)} (7.5,-.7);
\draw[decorate,decoration={brace,mirror},thick] (-.5,-1.5) -- node[below]{pseudo-spectrum} (10.5,-1.50);
\draw[->,very thick] (3,1.2) --  node[above,yshift=2mm]{last element in window} (3,.7);
\draw[->,very thick] (5,1.4) --  node[above,yshift=4.5mm]{CUT} (5,.7);

\draw[<-, thick] (8,-3.0) --  node[right,xshift=3mm, yshift=0mm]{label vector} (8.5,-3.0);

\end{tikzpicture}

%
%
%
%
}
\caption{Schematic of the sliding window and the corresponding label vector. Here the 1 indicates that the UE moves one grid point (discrete angle from grid) in the next time frame. }
\label{fig:sliding_wind}
\end{figure}
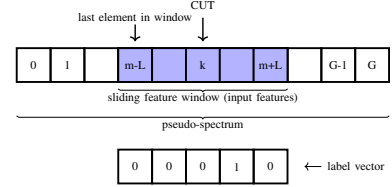

\begin{figure}
\vspace{-0.5cm}
\centering
\resizebox{1.25\linewidth}{!}{
%
%
\begin{tikzpicture}

\begin{axis}[%
width=8.422in,
height=1.66in,
at={(0.446in,0.643in)},
scale only axis,
axis on top,
xmin=0.5,
xmax=90.5,
xlabel style={font=\color{white!15!black}},
xlabel={Time Sequence Index},
y dir=reverse,
ymin=0.5,
ymax=17.5,
ylabel style={font=\color{white!15!black}},
ylabel={Feature Index},
label style={font=\huge},
axis background/.style={fill=white}
]
\addplot [forget plot] graphics [xmin=0.5, xmax=90.5, ymin=0.5, ymax=17.5] {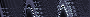};
\end{axis}

\begin{axis}[%
width=11.125in,
height=2.49in,
at={(0in,0in)},
scale only axis,
xmin=0,
xmax=1,
ymin=0,
ymax=1,
axis line style={draw=none},
ticks=none,
axis x line*=bottom,
axis y line*=left
]
\end{axis}


%

\end{tikzpicture}%
} 
\caption{Graphical representation of the input features for the RNN in a dynamic scenario where the UE has a cyclic motion. The window length parameter $L$ is set to 8 for a total of 90 time frames. }
\label{fig:input_feature}
\end{figure}

\noindent The input data sequence of interest are the observed signal values $z_k$ at frame $k$. These input values referred to as input features, are used to train the network to output estimates of the UE's future AoD under a supervised learning framework. The input features of the NN are generated based on the \textit{windowed} input method where features corresponding to the previous time steps are inputs to the current step. Due to the fact that the measurements in (\ref{measurements}) are equivalent to the beam space representation of the channel at discrete angles, they can be viewed as a \textit{pseudo-spectrum}. On this basis, we define our input features using a \textit{sliding window} technique consisting of these values.  As the initial feature set, we take the pseudo-spectrum $\mathcal{S} \equiv \zv$ described in section \ref{initial_spectrum} and define the window length parameter $L$ such that $L$ bins around the current main beam direction $\tilde{\theta}_m[k]$ are selected at each time step (total of $L'=2L+1$ bins, as in Fig. \ref{fig:sliding_wind}). These inputs are updated at every measurement frame. With reference to Fig.~\ref{fig:frame_strc}, the updates are made at interval $T_{\msubl{SP,FB}}$ which relates to the secondary probe transmitted during interval $T_{\msubl{SP}}$ of the previous frame. The number of directions to be scanned during the secondary probation can be adaptively selected, leading to a trade-off between overhead and improved channel exploration. If the user has moved out of span of the current window, at the next time instance (this can happen if for example the window length is very small or that the grid points are very fine) the initial BA shall be performed by using the initialization frame (a) and the label generated for the current time frame will be the last element of the label vector (i.e. farthest from current) to minimize the distance to the next user position (note that this will only happen during  inference mode). An exemplary graphical representation of this input is depicted in Fig. \ref{fig:input_feature}.

The labels generated for the supervised learning task consist of a \textit{ hot-one-encoded} vector with the same dimension as the sliding window such that the vector contains all zeros except at the position which corresponds to the next angular position (i.e. AoD) of the UE. The network outputs a value between '0' and '1' for each position in the window which can be interpreted as the probability of the target AoD in the next instant.  The significance of this technique is that the NN output becomes invariant to the value of the AoD itself, rather the predicted value is the amount of correction (in discrete grid points) needed for the next beam. This classification-type solution is an alternative to works attempting to regress values \cite{Guo}. It is well known that for limited data inputs and training, classification outperforms regression.  It's also worth noting that by defining the problem as a classification task, we directly bypass the problem of dealing with ambiguities within the regressed values (e.g. regressing $370^{\circ}$ and $10^{\circ}$ for a desired value of $10^{\circ}$).

\begin{algorithm}
  \algsetup{linenosize=\small}
  \scriptsize
\SetKwInOut{Parameter}{parameter}
\SetKwData{Left}{left}
\SetKwData{This}{this}
\SetKwData{Up}{up}
\SetKwFunction{Union}{Union}
\SetKwFunction{FindCompress}{FindCompress}
\SetKwInOut{Input}{input}
\SetKwInOut{Output}{output}
\Input{measurements $z_k$ and $\tilde{\theta}_m[k]$ as in (\ref{measurements})  for time frames $k \in \{0,1,..., K\}$ and sliding window size $2L+1$}
\Output{feature set $\mathcal{F}$, label set $\mathcal{L}$}
\BlankLine
 initialization\; set $\mathcal{W}=\mathcal{S}$ and $\mathcal{F}_k=[w_{k-L},....,w_{k+L}]$\
 \While{$k \leq K-1$}{
  insert $z(\tilde{\theta}_m[k])$ at $m^{th}$ position of $\mathcal{W}$ \\ 
  center sliding window of length 2$L+1$ at position $m$\\
  acquire $\tilde{\theta}_{m}[k+1]$ and calculate $d_{\tilde{\theta}} = \tilde{\theta}_{m}[k+1] - \tilde{\theta}_m[k] $ \;
  \eIf{$\vert d_{\tilde{\theta}} \vert\leq L $}{
   set position $(L+1+d_{\tilde{\theta}})$ of $\mathcal{L}_k$ to 1 and the rest 0\;
   }{
   set position $(L+1+ \text{sgn}(d_{\tilde{\theta}})L)$ of $\mathcal{L}_k$ to 1 and the rest 0\;
  }
 }
 \caption{Algorithm for feature and label generation from dataset. $m$ denotes the index of the main beam.}
 \label{alg: feature_gen}
\end{algorithm}

\subsection{Neural Network Architecture}
\begin{figure}[h]
\centering
\scalebox{.45}{\begin{tikzpicture}[
             font = \sffamily,
        shorten > = 1pt,
                > = Stealth,
    node distance = 6mm and 18mm,
      start chain = going below,
 every pin/.style = {pin distance=11mm, pin edge={thin, black, ->}},
    neuron/.style = {circle, draw, fill=#1,
                     minimum size=17pt, inner sep=1pt,
                     },
     annot/.style = {text width=6em, align=center},
        BC/.style = {decorate,
                            decoration={calligraphic brace, amplitude=4pt,
                            pre =moveto, pre  length=1pt,
                            post=moveto, post length=1pt,
                            raise=6mm, mirror},
                            thick,
                            pen colour={blue}
                         }
                        ]

\foreach \i in {1,...,9}
{
 \ifnum\i=5
    \node[neuron=white,draw=none,on chain]   (in-\i)        {$\vdots$};
 \else
     \ifnum\i<10
        \node[neuron=white,on chain]             (in-\i)    {};
     \else
        \node[neuron=white,draw=none,on chain]   (in-\i)    {$\vdots$};
     \fi
 \fi
}

\path   let \p1 = ($(in-1.north) - (in-9.south)$),
            \n1 = {veclen(\y1,\x1)} in
        node (r) [minimum height=\n1,
                  minimum width=9mm, draw, fill=gray!10,
                  below right=0mm and 15 mm of in-1.north]
                  {\rotatebox{90}{RNN (LSTM) Layer}};
\path   let \p2 = ($(in-1.north) - (in-9.south)$),
            \n2 = {veclen(\y2,\x2)} in
        node (r2) [minimum height=\n2,
                  minimum width=9mm, draw, fill=gray!10,
                  below right=0mm and 26 mm of in-1.north]
                  {\rotatebox{90}{RNN (biLSTM) Layer}};

\path   let \p2 = ($(in-2.north) - (in-7.south)$),
            \n2 = {veclen(\y2,\x2)} in
        node (r3) [minimum height=\n2,
                  minimum width=9mm, draw, fill=red!10,
                  below right=18.5mm and 50 mm of in-1.north]
                  {\rotatebox{90}{Dense Layer}};
                               
\path   let \p2 = ($(in-2.north) - (in-7.south)$),
            \n2 = {veclen(\y2,\x2)} in
        node (r4) [minimum height=\n2,
                  minimum width=9mm, draw, fill=green!10,
                  below right=18.5mm and 60 mm of in-1.north]
                  {\rotatebox{90}{Sigmoid Activation}};

\path   let \p2 = ($(in-2.north) - (in-7.south)$),
            \n2 = {veclen(\y2,\x2)} in
        node (out) [minimum height=\n2,
                  minimum width=9mm, draw, fill=blue!20,
                  below right=18.5mm and 80 mm of in-1.north]
                  {\rotatebox{90}{Prediction}};




\node[annot,above=of in-1]  {input layer};
\node[annot,above=of r]      {Recurrent Layer};
\node[annot,above=of r3]      {Fully connected layer};
\node[annot,above=of out]    {Classification layer};

\draw[BC] (in-1.north) --
    node[left=11mm,align=right] {Input features($\vert \vec{z}\vert$)\\ } (in-9.south) ;

\foreach \i in {1,...,4,6,7,8,9}
    {\draw[->]   (in-\i) -- (in-\i -| r.west);}



 \foreach \i in {2}
    \draw[->] (r2) edge (r3);

 \foreach \i in {2}
    \draw[->] (r4) edge (out);

   \end{tikzpicture}}
\caption{RNN architecture with the LSTM layer directly fed by the input data. Note that it is also possible to replace the sigmoid activation at the final layer with a softmax layer.}
\label{RNN_architecture}
\end{figure}
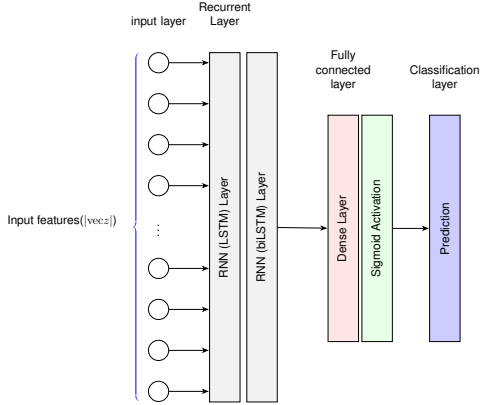

The recurrent network architecture is depicted in Fig.~\ref{RNN_architecture} consisting of two LSTM layers followed by a fully connected layer before the final classification layer. Here we have used a \textit{bi-directional LSTM} (biLSTM) layer in order to preserve the information contained within the time series from the future samples. For conciseness we refrain from discussing different recurrent implementations here. An important remark here is that as the final layer of the classifier we have chosen the sigmoid activation function as opposed to a softmax due to the fact that when angular grids of the labels are smaller than the beam width of the array, it is possible that a certain beam can cover multiple grid points.

\section{Simulations and Numerical Results} \label{simulations}

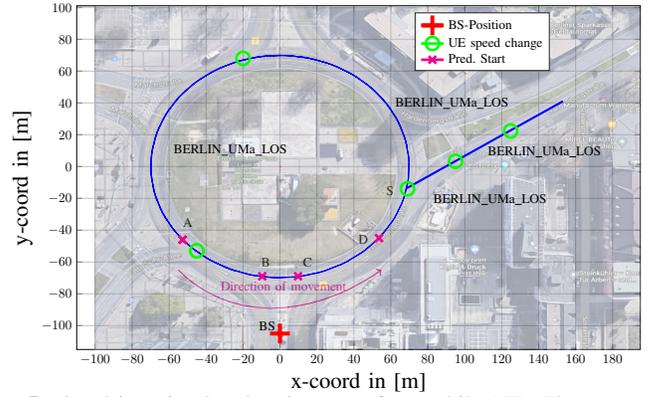
\begin{figure}[h]
\vspace{-5cm}
\centering
\scalebox{.5}{
%
%

%
%

\usetikzlibrary{decorations.markings}
\definecolor{mycolor1}{rgb}{0.30100,0.4500,0.93300}%
\definecolor{mycolor2}{rgb}{0.2400,0.18400,0.75600}%
\begin{tikzpicture}

\begin{axis}[%
width=5.9in,
height=3.6in,
at={(0.1in,0.756in)},
scale only axis,
xmin=-110,
xmax=195,
xlabel style={font=\color{white!15!black}},
xlabel={x-coord in [m]},
ymin=-115,
ymax=101,
ylabel style={font=\color{white!15!black}},
ylabel={y-coord in [m]},
axis background/.style={fill=white},
label style={font=\LARGE},
xmajorgrids,
ymajorgrids,
legend style={at={(0.6,0.811)}, anchor=south west, legend cell align=left, align=left, draw=white!15!black}
]
\addplot [color=red, line width=3.0pt, draw=none, mark size=7.5pt, mark=+, mark options={solid, red}]
  table[row sep=crcr]{%
0	-105\\
};
\addlegendentry{BS-Position}

\addplot [color=green, line width=2.0pt, draw=none, mark size=5.0pt, mark=o, mark options={solid, green}]
  table[row sep=crcr]{%
69	-14\\
};
\addlegendentry{UE speed change}

\addplot [color=magenta, line width=2.0pt, draw=none, mark size=4.0pt, mark=x, mark options={solid, magenta}]
  table[row sep=crcr]{%
-52.7	 -46\\
};
\addlegendentry{Pred. Start}

\addplot [color=green, line width=2.0pt, draw=none, mark size=5.0pt, mark=o, mark options={solid, green}]
  table[row sep=crcr]{%
-45	-53\\
};

\addplot [color=green, line width=2.0pt, draw=none, mark size=5.0pt, mark=o, mark options={solid, green}]
  table[row sep=crcr]{%
-20	 68\\
};

\addplot [color=magenta, line width=2.0pt, draw=none, mark size=4.0pt, mark=x, mark options={solid, magenta}]
  table[row sep=crcr]{%
-52.7	 -46\\
};

\addplot [color=magenta, line width=2.0pt, draw=none, mark size=4.0pt, mark=x, mark options={solid, magenta}]
  table[row sep=crcr]{%
-9.6 -69\\
};

\addplot [color=magenta, line width=2.0pt, draw=none, mark size=4.0pt, mark=x, mark options={solid, magenta}]
  table[row sep=crcr]{%
9.79  -69\\
};

\addplot [color=magenta, line width=2.0pt, draw=none, mark size=4.0pt, mark=x, mark options={solid, magenta}]
  table[row sep=crcr]{%
53.6	-45\\
};

\node[right, align=left]
at (axis cs:-55,-35) {A};

\node[right, align=left]
at (axis cs:-12.6,-60) {B};

\node[right, align=left]
at (axis cs:10,-60) {C};

\node[right, align=left]
at (axis cs:40,-46) {D};

\node[right, align=left, thick]
at (axis cs:55,-15) {S};

\coordinate (first) at (axis cs:-55,-65);
\coordinate (second) at (axis cs:55,-65);

\draw [->,thick,magenta] (first) to [out=-45,in=-145,looseness=1] (second);

\node[right, align=left, magenta,ultra thick]
at (axis cs:-35,-75) {Direction of movement};


\addplot [forget plot, opacity=.4]graphics [xmin=-100, xmax=200, ymin=-115, ymax=100] {ERP_bild_meas.png};

\addplot [color=blue,ultra thick]
  table[row sep=crcr]{%
68.035516357422	-14.00009200134277\\
153.149154663086	 41.04415092468262\\
};

\addplot [color=red, dashed, forget plot]
  table[row sep=crcr]{%
0	0\\
0	0\\
};
\node[left, align=right]
at (axis cs:0,-100) {BS};

\node[right, align=left]
at (axis cs:80.577,-20.141) {BERLIN\_UMa\_LOS};
\node[right, align=left]
at (axis cs:110,9.808) {BERLIN\_UMa\_LOS};

\addplot [color=green, line width=2.0pt, draw=none, mark size=5.0pt, mark=o, mark options={solid, green}, forget plot]
  table[row sep=crcr]{%
95 3.33197712898254\\
124.8477001190186	22.32169151306152\\
};

\addplot [color=blue, forget plot,thick]
  table[row sep=crcr]{%
68.7921013819068	12.9478487580545\\
65.2100917868156	25.4488492697623\\
59.3775913415809	37.07157464247\\
51.4958877384262	47.4149084785725\\
41.8369895170793	56.1218879595813\\
30.7342386722222	62.8920231288426\\
18.5708065582131	67.4916672173494\\
5.7664701133884	69.7620801182949\\
-7.2368752251859	69.6249067286635\\
-19.9904659470798	67.0848810926772\\
-32.0541579206161	62.2296630233541\\
-43.0116163782537	55.2268128405935\\
-52.4846842077414	46.3180086317801\\
-60.1464326793552	35.8107056052767\\
-65.7324442104939	24.0675253841002\\
-69.0499377837542	11.4937414299987\\
-69.9844220876313	-1.47670750667473\\
-68.5036467707169	-14.3961932160155\\
-64.6587154461256	-26.818846299887\\
-58.5823220348184	-38.3159437442018\\
-50.484171314067	-48.4907047456719\\
-40.6437417142916	-56.9919841684947\\
-29.4006401302069	-63.5263910507599\\
-17.1428816151411	-67.868413934092\\
-4.29349844255933	-69.8682035773337\\
8.70405932532399	-69.456744462012\\
21.4012279151827	-66.64823661375\\
33.3598104014915	-61.5396055396566\\
44.16709949112	-54.307157194439\\
53.4501206253055	-45.2004934170004\\
60.8885038497703	-34.5338978242611\\
66.2255402268666	-22.6754894469705\\
69.2770412176056	-10.0345184306062\\
69.9376952839923	2.95275775563137\\
68.1847023364888	15.8381301700806\\
64.0785605967757	28.1769067117977\\
57.760978720084	39.5432590626772\\
49.4499852326582	49.5449186142221\\
39.4324040644535	57.8367142021196\\
28.0539558568905	64.1324844425946\\
15.7073266632822	68.2149535592669\\
2.81861580765908	69.9432298720099\\
-10.1673693947928	69.2576681645421\\
-22.8024645560557	66.1819281236942\\
-34.6506149908507	60.8221578107504\\
-45.3029245695449	53.3633303444053\\
-54.3917672915781	44.0628602214928\\
-61.6034745694073	33.2417195851296\\
-66.6891603671757	21.2733610301973\\
-69.4733106035355	8.5708292355343\\
-69.8598403863989	-4.42749378168616\\
-67.8354100360354	-17.2730178383206\\
-63.4698854559488	-29.5224260556062\\
-56.9139269634973	-40.7529743404539\\
-48.393789793061	-50.5770808713297\\
-38.2035157136153	-58.6557020853007\\
-26.6947852388331	-64.710033542354\\
-14.2647806440429	-68.5311318539052\\
-1.34247865418588	-69.9871256093794\\
11.6261541388488	-69.0277664417713\\
24.193552202796	-65.6861631685897\\
35.9259971732427	-60.0776391605742\\
46.4185860768557	-52.3957523719771\\
55.3092050954379	-42.9056153866925\\
62.2910266168028	-31.9347460147214\\
67.1230982818807	-19.8617641975979\\
69.6386585853353	-7.10332530826934\\
69.7508920467588	5.90025920459404\\
67.4559253334953	18.7002175763254\\
62.832960934731	30.854805463243\\
56.0415437735734	41.9445511535724\\
47.3160550901329	51.5867321189035\\
36.9576236193495	59.4485833002814\\
25.3237332200572	65.2587812926304\\
12.815885610437	68.8168080923563\\
-0.134256013996861	69.9998712521866\\
-13.0797642768943	68.7671416190965\\
-25.5738717056477	65.1611624050943\\
-37.1853892974711	59.306380961879\\
-47.5135874508469	51.4048539298647\\
-56.2020257007218	41.7292739828458\\
-62.9508539740841	30.6135588250294\\
-67.5271608325014	18.4413272272224\\
-69.7730115693947	5.63265981017084\\
-69.6108987561489	-7.37039852119215\\
-67.046417131138	-20.1190941614538\\
-62.1680705175329	-32.1734519150038\\
-55.1442174334235	-43.117459151199\\
-46.2172608054952	-52.5734229781249\\
-35.6952823073095	-60.2150049489451\\
-23.9414100327968	-65.7784834550136\\
};
\node[right, align=left]
at (axis cs:-60,10.8) {BERLIN\_UMa\_LOS};
\node[right, align=left]
at (axis cs:60,40.229) {BERLIN\_UMa\_LOS};

\end{axis}

\begin{axis}[%
width=10.656in,
height=8.417in,
at={(0in,0in)},
scale only axis,
xmin=0,
xmax=1,
ymin=0,
ymax=1,
axis line style={draw=none},
ticks=none,
axis x line*=bottom,
axis y line*=left,
legend style={legend cell align=left, align=left, draw=white!15!black}
]
\end{axis}
\end{tikzpicture}%

}
\vspace{-1cm}
\hspace{1.5cm}
\caption{Quadriga simulated trajectory of a mobile UE. The green markings on the trajectory imply that the linear speed of the UE changes from there onward. The notation BERLIN-UMa-LOS describes the scenario defined by the Quadriga software to generate the coefficients (in this case a LoS environment in Berlin, Germany).}
\label{quadriga_traj}
\end{figure}

In this section we use the Fraunhofer Quadriga \cite{quadriga} channel generator with a mobile UE to train a network to predict the UE's AoD. The parameters generated by Quadriga at each time instance include channel coefficients, delay values and position coordinates (trajectory) among others. In the following we assume a ULA at the BS with N$_{\msubl{TX}}=64$. Since this ULA is not available in Quadriga, we have self-defined a 64 element ULA to obtain the channel parameters. These values are then used to simultaneously generate multiple realizations of the measurement values for the NN and state variables for the PF along the same trajectory. As can be seen in Fig.~\ref{quadriga_traj}, the UE trails a trajectory with variable speed along a circular path (start:S) followed by a linear path. The circular path can be a representation of a likely scenario for a BS installed at the corner of an urban round-about. The significant challenge of a circular path is the non-linearity in AoD values with time. The dataset contains 26 realizations of the the same trajectory, where in each case the parameters vary.  

\begin{figure*}[t]  
\centering 
\begin{subfigure}[h]{0.3\linewidth}
\scalebox{0.3}{
%
%
\definecolor{mycolor1}{rgb}{0.00000,1.00000,1.00000}%

\begin{tikzpicture}

\begin{axis}[%
width=7.0582in,
height=5.05066in,
at={(0.003in,2.81in)},
scale only axis,
name=plot1,
xmin=0,
xmax=43.8128631536766,
xlabel style={font=\color{white!15!black}},
xlabel={Time Instance (segment A-B on prediction trajectory)},
ymin=-45.932944606414,
ymax=-15.32944606414,
xtick = {0,5,...,45},
ytick = {-45,-40,...,-15},
ylabel style={font=\huge, font=\color{white!25!black}},
ylabel={Angle of Departure (deg$^{\circ}$)},
label style={font=\huge},
tick label style={font=\huge}, 
axis background/.style={fill=white},
legend style={at={(0.65,0.901)}, legend cell align=left, align=left, draw=white!15!black,font=\huge}
]
\addplot [color=blue, dashdotted, mark=o,  thick, mark size=3.5, mark options={solid, blue}]
  table[row sep=crcr]{%
1	-41.7879769904096\\
2	-41.8063663527335\\
3	-41.8096833245242\\
4	-41.7972073663874\\
5	-41.7681764817041\\
6	-41.7217846805301\\
7	-41.6571793201236\\
8	-41.5734583266407\\
9	-41.4696673059137\\
10	-41.3447965554984\\
11	-41.1977779955253\\
12	-41.0274820425439\\
13	-40.8327144587707\\
14	-40.61221321925\\
15	-40.3646454517459\\
16	-40.0886045191199\\
17	-39.782607331934\\
18	-39.4450920005389\\
19	-39.0744159614592\\
20	-38.6688547429795\\
21	-38.2266015699306\\
22	-37.7457680481506\\
23	-37.2243862151622\\
24	-36.6604122951729\\
25	-36.0517325530595\\
26	-35.3961717023607\\
27	-34.6915043844147\\
28	-33.9354702963315\\
29	-33.125793599597\\
30	-32.2602072818607\\
31	-31.3364831625586\\
32	-30.3524682164291\\
33	-29.3061278227897\\
34	-28.1955964151506\\
35	-27.0192357860001\\
36	-25.7757009759295\\
37	-24.4640132277967\\
38	-23.083638905177\\
39	-21.6345725618139\\
40	-20.1174215255577\\
41	-18.5334884716838\\
42	-16.8848475819687\\
};
\addlegendentry{Oracle}

\addplot [color=cyan, draw=none, mark=square, thick,mark size=2.5,mark options={solid, cyan}]
  table[row sep=crcr]{%
1	-42\\
2	-42\\
3	-42\\
4	-42\\
5	-42\\
6	-42\\
7	-42\\
8	-42\\
9	-41\\
10	-41\\
11	-41\\
12	-41\\
13	-41\\
14	-41\\
15	-40\\
16	-40\\
17	-40\\
18	-39\\
19	-39\\
20	-39\\
21	-38\\
22	-38\\
23	-37\\
24	-37\\
25	-36\\
26	-35\\
27	-35\\
28	-34\\
29	-33\\
30	-32\\
31	-31\\
32	-30\\
33	-29\\
34	-28\\
35	-27\\
36	-26\\
37	-24\\
38	-23\\
39	-22\\
40	-20\\
41	-19\\
42	-17\\
};
\addlegendentry{GT}

\addplot [color=red, draw=none, mark=x,thick, mark size=4.5,mark options={solid, red}]
  table[row sep=crcr]{%
1	-43\\
2	-42\\
3	-42\\
4	-42\\
5	-42\\
6	-42\\
7	-42\\
8	-42\\
9	-41\\
10	-41\\
11	-41\\
12	-41\\
13	-41\\
14	-41\\
15	-40\\
16	-40\\
17	-40\\
18	-39\\
19	-39\\
20	-39\\
21	-38\\
22	-38\\
23	-37\\
24	-37\\
25	-36\\
26	-35\\
27	-35\\
28	-34\\
29	-33\\
30	-32\\
31	-31\\
32	-30\\
33	-29\\
34	-28\\
35	-27\\
36	-26\\
37	-24\\
38	-22\\
39	-22\\
40	-20\\
41	-20\\
42	-17\\
};
\addlegendentry{RNN}

\addplot [color=green, mark=x,thick,mark size=3.5, mark options={solid, green}]
  table[row sep=crcr]{%
1	-40.6688852291452\\
2	-40.973985252883\\
3	-41.2805646038457\\
4	-41.4929954822708\\
5	-40.9669506423078\\
6	-41.4177909037093\\
7	-41.1115899225185\\
8	-42.8137368724835\\
9	-42.7059689342888\\
10	-42.2245552861141\\
11	-40.3513541078583\\
12	-40.2573881232319\\
13	-39.6140395949038\\
14	-40.0192720111306\\
15	-40.3489592188952\\
16	-41.0212030041058\\
17	-40.9241595059776\\
18	-40.2434151437957\\
19	-38.0816495528988\\
20	-37.8409735138191\\
21	-39.5154122953394\\
22	-39.3273394133752\\
23	-38.3423256005105\\
24	-37.6579222089349\\
25	-37.3469698475328\\
26	-36.5330273613935\\
27	-35.5559975595095\\
28	-34.3280619367746\\
29	-33.8420278486914\\
30	-33.8423511519569\\
31	-32.6167648342206\\
32	-31.1436890514599\\
33	-30.4290257687891\\
34	-28.426853751496\\
35	-27.8621539675105\\
36	-27.0057933383601\\
37	-26.2771717125203\\
38	-24.3805707801567\\
39	-23.4146832531551\\
40	-20.5511301141739\\
41	-20.3039790779177\\
42	-18.2900460240438\\
};
\addlegendentry{PF}
\end{axis}

%
%

\begin{axis}[%
width=13.292in,
height=4.375in,
at={(0in,0in)},
scale only axis,
xmin=0,
xmax=1,
ymin=0,
ymax=1,
axis line style={draw=none},
ticks=none,
axis x line*=bottom,
axis y line*=left,
legend style={legend cell align=left, align=left, draw=white!15!black}
]
\end{axis}
\end{tikzpicture}%
}
\end{subfigure}
\begin{subfigure}[h]{0.3\linewidth}
\hspace{0.5cm}
\scalebox{0.3}{
%
%
\definecolor{mycolor1}{rgb}{0.00000,1.00000,1.00000}%

\begin{tikzpicture}

\begin{axis}[%
width=5.582in,
height=5.05066in,
at={(0.003in,2.81in)},
scale only axis,
name=plot1,
xmin=42,
xmax=60,
xlabel style={font=\color{white!15!black}},
xlabel={Time Instance (segment  B-C)},
ymin=-20.932944606414,
ymax=20.32944606414,
label style={font=\huge},
tick label style={font=\huge}, 
axis background/.style={fill=white},
legend style={at={(0.85,0.901)}, legend cell align=left, align=left, draw=white!15!black}
]
\addplot [color=blue, dashdotted, mark=o, thick, mark size=3.5, mark options={solid, blue}]
  table[row sep=crcr]{%
43	-15.1744091244358\\
44	-13.4059667781791\\
45	-11.584221915813\\
46	-9.71477948172688\\
47	-7.80411116686414\\
48	-5.85948330642976\\
49	-3.88884924027383\\
50	-1.90070858619226\\
51	0.096061310582235\\
52	2.0923939327254\\
53	4.07923239656452\\
54	6.04772773210706\\
55	7.98942591412822\\
56	9.89643404009663\\
57	11.7615580602223\\
58	13.5784070616165\\
59	15.3414619267369\\
};

\addplot [color=cyan, thick,draw=none, mark=square,mark size=2.5,  mark options={solid, cyan}]
  table[row sep=crcr]{%
43	-15\\
44	-13\\
45	-12\\
46	-10\\
47	-8\\
48	-6\\
49	-4\\
50	-2\\
51	0\\
52	2\\
53	4\\
54	6\\
55	8\\
56	10\\
57	12\\
58	14\\
59	15\\
};

\addplot [color=red, draw=none, thick,mark=x,mark size=4.5,  mark options={solid, red}]
  table[row sep=crcr]{%
43	-15\\
44	-13\\
45	-12\\
46	-10\\
47	-7\\
48	-6\\
49	-3\\
50	-2\\
51	0\\
52	2\\
53	5\\
54	6\\
55	8\\
56	10\\
57	11\\
58	14\\
59	15\\
};

\addplot [color=green, thick,mark=x,mark size=3.5,  mark options={solid, green}]
  table[row sep=crcr]{%
43	-16.0114051343287\\
44	-14.2709666767958\\
45	-12.742524330539\\
46	-9.0507794681729\\
47	-8.62133703408683\\
48	-6.65066871922407\\
49	-5.5360408587897\\
50	-2.22540679263377\\
51	-0.78665717632018\\
52	1.080496241777704\\
53	2.64583638036546\\
54	4.32267484420458\\
55	8.38117017974713\\
56	8.992554191053581\\
57	11.1098764877367\\
58	12.8750005078624\\
59	14.6018495092566\\
};

\end{axis}

%
%

\begin{axis}[%
width=13.292in,
height=4.375in,
at={(0in,0in)},
scale only axis,
xmin=0,
xmax=1,
ymin=0,
ymax=1,
axis line style={draw=none},
ticks=none,
axis x line*=bottom,
axis y line*=left,
legend style={legend cell align=left, align=left, draw=white!15!black}
]
\end{axis}
\end{tikzpicture}%
}
\end{subfigure}
\begin{subfigure}[h]{0.3\linewidth}
\scalebox{0.3}{
%
%
\definecolor{mycolor1}{rgb}{0.00000,1.00000,1.00000}%

\begin{tikzpicture}

\begin{axis}[%
width=7.0582in,
height=5.05066in,
at={(0.003in,2.81in)},
scale only axis,
name=plot1,
xmin=59,
xmax=99.8128631536766,
xlabel style={font=\color{white!15!black}},
xlabel={Time Instance (segment  C-D)},
ymin=14.032944606414,
ymax=44.32944606414,
xtick = {60,65,...,100},
ytick = {15,20,...,45},
label style={font=\huge},
tick label style={font=\huge}, 
axis background/.style={fill=white},
legend style={at={(0.85,0.901)}, legend cell align=left, align=left, draw=white!15!black}
]
\addplot [color=blue, dashdotted,mark size=3.5,  thick,mark=o, mark options={solid, blue}]
 table[row sep=crcr]{%
60	17.0461088699582\\
61	18.6886406193824\\
62	20.2662296744386\\
63	21.7768790685589\\
64	23.219356434974\\
65	24.5931170170886\\
66	25.8982207239297\\
67	27.1352475529714\\
68	28.3052148190027\\
69	29.4094987426197\\
70	30.4497621378132\\
71	31.4278892376485\\
72	32.345928128529\\
73	33.2060408274568\\
74	34.0104607220226\\
75	34.7614568822941\\
76	35.4613046277457\\
77	36.1122616716831\\
78	36.716549152954\\
79	37.2763368854962\\
80	37.7937321986741\\
81	38.2707717963487\\
82	38.7094161235112\\
83	39.1115457913784\\
84	39.4789596719151\\
85	39.8133743288657\\
86	40.1164245034256\\
87	40.3896644182102\\
88	40.634569703127\\
89	40.8525397813498\\
90	41.0449005832299\\
91	41.2129074811212\\
92	41.3577483592535\\
93	41.4805467504627\\
94	41.5823649862488\\
95	41.6642073187148\\
96	41.7270229828432\\
97	41.7717091756227\\
98	41.7991139350591\\
};

\addplot [color=cyan, thick,draw=none, mark=square,mark size=2.5,  mark options={solid, cyan}]
  table[row sep=crcr]{%
60	17\\
61	19\\
62	20\\
63	22\\
64	23\\
65	25\\
66	26\\
67	27\\
68	28\\
69	29\\
70	30\\
71	31\\
72	32\\
73	33\\
74	34\\
75	35\\
76	35\\
77	36\\
78	37\\
79	37\\
80	38\\
81	38\\
82	39\\
83	39\\
84	39\\
85	40\\
86	40\\
87	40\\
88	41\\
89	41\\
90	41\\
91	41\\
92	41\\
93	41\\
94	42\\
95	42\\
96	42\\
97	42\\
98	42\\
};

\addplot [color=red, draw=none,thick, mark=x,mark size=4.5,  mark options={solid, red}]
  table[row sep=crcr]{%
60	17\\
61	19\\
62	20\\
63	22\\
64	23\\
65	25\\
66	25\\
67	27\\
68	28\\
69	29\\
70	30\\
71	31\\
72	32\\
73	33\\
74	34\\
75	35\\
76	36\\
77	36\\
78	37\\
79	37\\
80	38\\
81	38\\
82	39\\
83	39\\
84	39\\
85	40\\
86	40\\
87	40\\
88	41\\
89	41\\
90	41\\
91	41\\
92	41\\
93	41\\
94	42\\
95	42\\
96	42\\
97	42\\
98	42\\
};

\addplot [color=green, mark=x,thick, mark size=3.5, mark options={solid, green}]
  table[row sep=crcr]{%
60	15.454904374377\\
61	17.1095513175982\\
62	19.4520830670224\\
63	21.5244835939006\\
64	22.7024426990708\\
65	24.42798882614\\
66	24.3765594647288\\
67	26.4716631715698\\
68	27.0086900006115\\
69	28.132053059549\\
70	29.7729411902598\\
71	30.0132045854533\\
72	32.813316852885\\
73	33.0093705761691\\
74	33.8594832750968\\
75	34.9439031696627\\
76	35.0048993299341\\
77	35.3547470753858\\
78	35.5139705672618\\
79	36.7899916005941\\
80	37.0849723757133\\
81	37.0271746463142\\
82	37.8942142439887\\
83	39.8228585711514\\
84	39.2049882390184\\
85	40.1478746577279\\
86	39.7168167765058\\
87	39.6440614251437\\
88	39.9031068658503\\
89	39.9935336260478\\
90	40.8659822289898\\
91	42.00834303087\\
92	42.173465292672\\
93	42.5410090529209\\
94	42.6439891981027\\
95	42.5124086279227\\
96	42.6590848777316\\
97	42.3904654304833\\
98	42.6045166565952\\
};
\end{axis}

%
%

\begin{axis}[%
width=13.292in,
height=4.375in,
at={(0in,0in)},
scale only axis,
xmin=0,
xmax=1,
ymin=0,
ymax=1,
axis line style={draw=none},
ticks=none,
axis x line*=bottom,
axis y line*=left,
legend style={legend cell align=left, align=left, draw=white!15!black}
]
\end{axis}
\end{tikzpicture}%
}
\end{subfigure}
\vspace*{-1.0cm}
\caption{The performance of PF compared to the proposed method. The overall MSE is reported in Tab.\ref{table_mse} for other training window lengths and transmitted number of secondary probes. The results are segmented in multiple plots for better visualization}
\label{prediction}
\end{figure*}
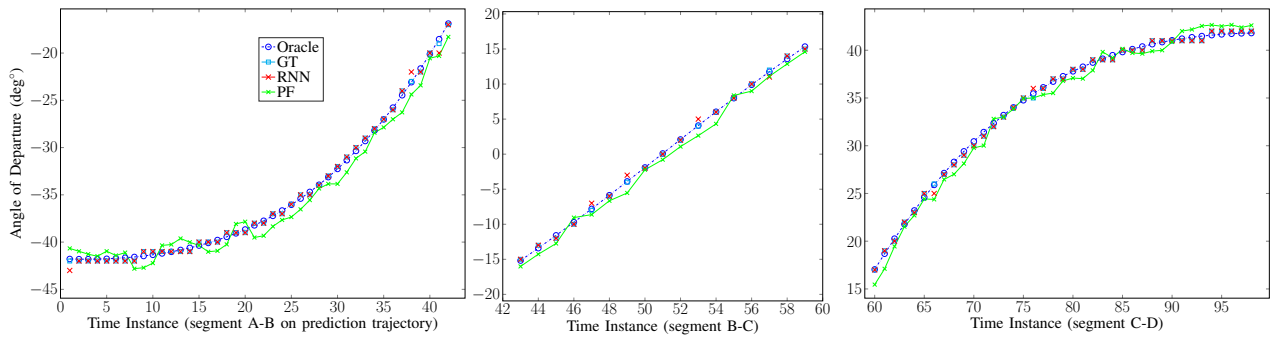

Fig.~\ref{prediction} shows the performance of PF compared to the proposed method. The number of particles is set to 100 as in \cite{PF} and the other parameters have been accommodated correspondingly to the array size. The \textit{Oracle} defines the real AoD, \textit{GT} denotes the Ground Truth value which is the rounded (within 1 degree) value of the corresponding real value, included in the plots to signify that the network has been trained to predict up to a pre-chosen \textit{gridded} accuracy. 
Fig. \ref{prediction}  shows the estimated AoD values associated with the UE along the marked circular section (A-D) of the trajectory from Fig.~\ref{quadriga_traj}. A few remarks are in order. 1) The section of the path used for inference, has not been seen by the network during training and validation. 2) Note that when the AoD progression becomes highly nonlinear along the path, the PF deviates significantly  from the true value. This could be attributed to the channel model considered in \cite{PF}, which ignores the Doppler induced time-variance of the channel. Additionally, even though the PF outperforms other EKF variants, the particles are not fully able to capture the channel dynamics. For both estimators, at \textit{each} time index the reported values are the average estimated value (binned avg. for RNN) over 26 independent runs. In Tab.~\ref{table_mse}, MSE of the RNN estimated angles for various window lengths ($L'$) and probe lengths are shown. It is observed that with a larger window (more memory) and larger number of probes, the estimation error decreases.

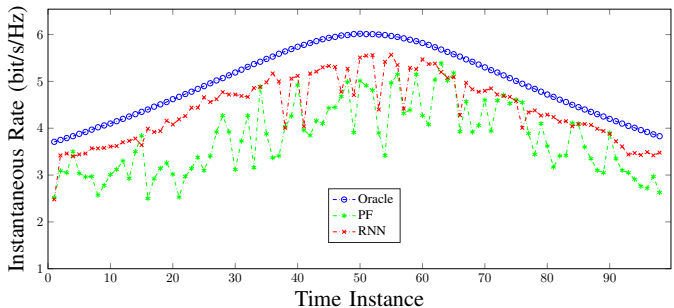
\begin{figure}[h]
\centering
\scalebox{0.5}{
%
%

%
%

\definecolor{mycolor1}{rgb}{0.00000,1.00000,1.00000}%
\begin{tikzpicture}

\begin{axis}[%
width=6.521in,
height=2.719in,
at={(0.0in,0.448in)},
scale only axis,
xmin=0,
xmax=99.8211091234347,
xlabel style={font=\color{white!15!black}},
xlabel={Time Instance},
ymin=1,
ymax=6.54263322884013,
ylabel style={font=\color{white!20!black}},
ylabel={Instantaneous Rate (bit/s/Hz)},
label style={font=\LARGE},
axis background/.style={fill=white},
legend style={at={(0.45,0.100)}, anchor=south west, legend cell align=left, align=left, draw=white!15!black}
]
\addplot [color=blue, dashdotted, mark=o, mark options={solid, blue}]
  table[row sep=crcr]{%
1	3.71\\
2	3.75\\
3	3.79\\
4	3.83\\
5	3.88\\
6	3.92\\
7	3.970\\
8	4.02\\
9	4.06\\
10	4.10\\
11	4.15\\
12	4.20\\
13	4.25\\
14	4.30\\
15	4.35\\
16	4.40\\
17	4.460\\
18	4.51\\
19	4.56\\
20	4.62\\
21	4.67\\
22	4.73\\
23	4.78\\
24	4.84\\
25	4.90\\
26	4.96\\
27	5.02\\
28	5.07\\
29	5.13\\
30	5.19\\
31	5.25\\
32	5.31\\
33	5.37\\
34	5.42\\
35	5.48\\
36	5.53\\
37	5.59\\
38	5.64\\
39	5.69\\
40	5.74\\
41	5.78\\
42	5.82\\
43	5.86\\
44	5.90\\
45	5.93\\
46	5.95\\
47	5.980\\
48	5.99\\
49	6.01\\
50	6.017\\
51	6.01\\
52	6.01\\
53	6.00\\
54	5.99\\
55	5.97\\
56	5.95\\
57	5.92\\
58	5.89\\
59	5.86\\
60	5.82\\
61	5.780\\
62	5.73\\
63	5.68\\
64	5.63\\
65	5.58\\
66	5.53\\
67	5.47\\
68	5.42\\
69	5.36\\
70	5.30\\
71	5.24\\
72	5.190\\
73	5.13\\
74	5.07\\
75	5.01\\
76	4.95\\
77	4.89\\
78	4.84\\
79	4.78\\
80	4.72\\
81	4.67\\
82	4.61\\
83	4.56\\
84	4.50\\
85	4.45\\
86	4.40\\
87	4.35\\
88	4.30\\
89	4.250\\
90	4.20\\
91	4.15\\
92	4.10\\
93	4.05\\
94	4.01\\
95	3.96\\
96	3.92\\
97	3.87\\
98	3.83\\
};
\addlegendentry{Oracle}

\addplot [color=green, dashdotted, mark=asterisk, mark options={solid, green}]
  table[row sep=crcr]{%
1	2.52\\
2	3.09\\
3	3.05\\
4	3.50\\
5	3.04\\
6	2.96\\
7	2.97\\
8	2.57\\
9	2.78\\
10	3.01\\
11	3.12\\
12	3.30\\
13	2.93\\
14	3.50\\
15	3.84\\
16	2.50\\
17	2.920\\
18	3.14\\
19	3.26\\
20	3.02\\
21	2.53\\
22	2.97\\
23	3.14\\
24	3.380\\
25	3.10\\
26	3.410\\
27	3.92\\
28	4.27\\
29	3.92\\
30	3.12\\
31	3.73\\
32	4.27\\
33	3.16\\
34	4.89\\
35	3.88\\
36	3.37\\
37	3.410\\
38	4.03\\
39	4.26\\
40	4.92\\
41	3.96\\
42	3.850\\
43	4.16\\
44	4.100\\
45	4.43\\
46	4.45\\
47	4.680\\
48	4.99\\
49	3.910\\
50	5.01\\
51	4.91\\
52	4.81\\
53	3.90\\
54	3.42\\
55	4.97\\
56	5.15\\
57	4.32\\
58	4.39\\
59	5.15\\
60	4.27\\
61	4.080\\
62	5.03\\
63	5.39\\
64	5.02\\
65	5.18\\
66	3.93\\
67	4.57\\
68	3.92\\
69	4.06\\
70	4.60\\
71	3.94\\
72	4.590\\
73	4.71\\
74	4.53\\
75	4.61\\
76	4.55\\
77	3.89\\
78	3.44\\
79	4.10\\
80	3.62\\
81	3.17\\
82	3.41\\
83	3.42\\
84	4.11\\
85	4.09\\
86	3.60\\
87	3.35\\
88	3.100\\
89	3.050\\
90	3.900\\
91	3.35\\
92	3.10\\
93	3.05\\
94	2.91\\
95	2.76\\
96	2.72\\
97	2.97\\
98	2.63\\
};
\addlegendentry{PF}

\addplot [color=red, dashdotted, mark=x, mark options={solid, red}]
  table[row sep=crcr]{%
1	2.480\\
2	3.42\\
3	3.46\\
4	3.40\\
5	3.44\\
6	3.46\\
7	3.57\\
8	3.57\\
9	3.58\\
10	3.61\\
11	3.62\\
12	3.70\\
13	3.73\\
14	3.78\\
15	3.64\\
16	3.990\\
17	3.920\\
18	3.94\\
19	4.16\\
20	4.08\\
21	4.19\\
22	4.26\\
23	4.44\\
24	4.440\\
25	4.66\\
26	4.56\\
27	4.62\\
28	4.77\\
29	4.72\\
30	4.72\\
31	4.69\\
32	4.67\\
33	4.86\\
34	4.88\\
35	4.98\\
36	5.17\\
37	5.000\\
38	4.01\\
39	5.06\\
40	5.12\\
41	4.04\\
42	5.170\\
43	5.21\\
44	5.300\\
45	5.33\\
46	5.31\\
47	4.780\\
48	5.27\\
49	4.710\\
50	5.51\\
51	5.55\\
52	5.56\\
53	4.40\\
54	5.42\\
55	5.57\\
56	5.35\\
57	4.42\\
58	5.29\\
59	5.26\\
60	5.470\\
61	5.37\\
62	5.380\\
63	5.20\\
64	5.08\\
65	5.09\\
66	4.28\\
67	4.97\\
68	4.83\\
69	4.78\\
70	4.80\\
71	4.85\\
72	4.72\\
73	4.69\\
74	4.670\\
75	4.58\\
76	4.01\\
77	4.340\\
78	4.38\\
79	4.27\\
80	4.29\\
81	4.24\\
82	4.14\\
83	4.15\\
84	4.04\\
85	4.09\\
86	4.09\\
87	4.070\\
88	3.99\\
89	3.94\\
90	3.88\\
91	3.72\\
92	3.61\\
93	3.44\\
94	3.47\\
95	3.43\\
96	3.49\\
97	3.42\\
98	3.480\\
};
\addlegendentry{RNN}

\end{axis}

\begin{axis}[%
width=5.833in,
height=4.375in,
at={(0in,0in)},
scale only axis,
xmin=0,
xmax=1,
ymin=0,
ymax=1,
axis line style={draw=none},
ticks=none,
axis x line*=bottom,
axis y line*=left,
legend style={legend cell align=left, align=left, draw=white!15!black}
]
\end{axis}
\end{tikzpicture}%

}
\caption{Comparison of achievable instantaneous rate.}
\label{rate_comparison}
\end{figure}

In Fig.~\ref{rate_comparison}, a comparison of the achievable rate between the proposed RNN scheme, the PF from \cite{PF} and an Oracle estimator is provided. The reported values correspond to the predictions form Fig.~\ref{prediction} . The plotted values at each time index are averaged for 26 runs. It's worth mentioning, the change in the Oracle rate occurs as the result of the UE moving closer to the BS, leading to a higher received signal power. The achievable rate is calculated according to: 
\begin{equation}
\begin{split} 
r(\tilde{\theta},\theta)&=\left(1-\frac{T_{\msubl{X}}}{T_{\msubl{FR}}}\right)
\cdot \log_2 \left(1+\frac{P\vert \cv^{H}(\tilde{\theta})\matr{H}(\theta)\fv(\tilde{\theta})\vert^{2}}{\sigma_{\eta}^2}\right)\\
\end{split} 
\label{eq: rate}
\end{equation}
where $\theta$ and $\tilde{\theta}$ are the actual and estimated AoD of the UE. The channel is defined $\matr{H(\theta)}=\alpha \av_{\msub{RX}}(\theta)\av_{\msub{TX}}^{H}(\theta)e^{j2\pi\nu T_s}$, where $T_s$ is the sampling time. $ T_{\msubl{X}}$ and $T_{\msubl{FR}} \in [T_F, T_{F'}]$ denote the intervals in each frame type of Fig.~\ref{fig:frame_strc} defined as:
\begin{equation*}
T_{\msubl{X}}= \left\{
    \begin{array}{ll}
        T_{\msubl{ACQ}}+T_{\msubl{FB,I}}+T_{\msubl{SP}} & \text{initialization frame} \\ 
        T_{\msubl{FD,SP}}+T_{\msubl{SP}} & \text{secondary frame}
    \end{array}
\right.
\end{equation*}
Note that the values for the proposed estimator in Fig.~\ref{rate_comparison} are obtained with a combination of $T_{\msubl{X}}$, depending on whether any future values move out of the current frame ( in Fig.~\ref{prediction} this is not the case, only a secondary $T_{\msubl{X}}$ is used). However for the Oracle we use the initial $T_{\msubl{X}}$ since a full BA must takes place for the exact AoD to be discovered. The above results have been obtained with $\sigma_{\eta}=1$ and we assume $P=1$.   

\vspace{-1.7cm}

\begin{table} 
\centering
\caption{RNN Estimation MSE vs. Window/Probe Length }  \label{table_mse}
  \begin{tabular}{lSSSS}
    \toprule
    \multirow{2}{*}{} &
      \multicolumn{2}{c}{\# probes $=1$} &
      \multicolumn{2}{c}{\# probes $=3$} \\
      & {$L'$=17} & {$L'$=29} & {$L'$=17} & {$L'$=29} \\
      \midrule
    MSE($^{\circ}$) &0.058&0.051&0.048&0.044 \\
    \bottomrule
  \end{tabular}
\end{table}

\section{Conclusion}
In this paper, we proposed a RNN based approach for beam tracking at the BS side. This model is trained based on past channel measurements to predict the amount on alignment correction for the next time frame. Furthermore, this model places no restriction on the mobility on the UE. Based on this model, we proposed a frame structure which can adapted be to channel conditions. 
Our simulations demonstrate that our proposed scheme outperforms PF both in terms of prediction and communication rates,  especially at the high mobility scenarios. 



\vspace{-0.5cm}
\begin{thebibliography}{00}

\bibitem{6G}
T. S. Rappaport, Y. Xing, O. Kanhere, S. Ju, A. Madanayake, S. Mandal, A. Alkhateeb, and G. C. Trichopoulos,
“Wireless Communications and Applications Above 100 GHz: Opportunities and Challenges for 6G and Beyond,”
IEEE Access, vol. 7, pp. 78 729–78 757, 2019.

\bibitem{X.Song_OSPS} 
X. Song, T. Kühne and G. Caire, "Fully-/Partially-Connected Hybrid Beamforming Architectures for mmWave MU-MIMO," in IEEE Transactions on Wireless Communications, vol. 19, no. 3, pp. 1754-1769, March 2020, doi: 10.1109/TWC.2019.2957227.

\bibitem{BeamTracking} 
Q. Duan, T. Kim, H. Huang, K. Liu and G. Wang, "AoD and AoA tracking with directional sounding beam design for millimeter wave MIMO systems," 2015 IEEE 26th Annual International Symposium on Personal, Indoor, and Mobile Radio Communications (PIMRC), Hong Kong, 2015, pp. 2271-2276, doi: 10.1109/PIMRC.2015.7343676.  

\bibitem{Va} 
V. Va, H. Vikalo, and R. W. Heath, “Beam tracking for mobile millimeter wave communication systems,” in 2016 IEEE Global Conference on Signal and Information Processing (GlobalSIP). IEEE, 2016, pp. 743–747.

\bibitem{Love}
S. G. Larew and D. J. Love, “Adaptive beam tracking with the unscented kalman filter for millimeter wave communication,” IEEE Signal Processing Letters, vol. 26, no. 11, pp. 1658–1662, 2019.

\bibitem{PF}
J. Lim, H. Park and D. Hong, "Beam Tracking Under Highly Nonlinear Mobile Millimeter-Wave Channel," in IEEE Communications Letters, vol. 23, no. 3, pp. 450-453, March 2019

\bibitem{Palacios}
J. Palacios, D. De Donno, and J. Widmer, “Tracking mm-wave channel dynamics: Fast beam training strategies under mobility,” in IEEE INFOCOM 2017-IEEE Conference on Computer Communications. IEEE, 2017, pp. 1–9.

\bibitem{Lim}Y. -G. Lim, Y. J. Cho, M. S. Sim, Y. Kim, C. -B. Chae and R. A. Valenzuela, "Map-Based Millimeter-Wave Channel Models: An Overview, Data for B5G Evaluation and Machine Learning," in IEEE Wireless Communications, vol. 27, no. 4, pp. 54-62, August 2020.

\bibitem{Guo}
Y. Guo, Z. Wang, M. Li and Q. Liu, "Machine Learning Based mmWave Channel Tracking in Vehicular Scenario," 2019 IEEE International Conference on Communications Workshops , 2019, pp. 1-6

\bibitem{Qlearning} 
H. -L. Chiang, K. -C. Chen, W. Rave, M. K. Marandi and G. Fettweis, "Machine-Learning Beam Tracking and Weight Optimization for mmWave Multi-UAV Links," in IEEE Transactions on Wireless Communications, doi: 10.1109/TWC.2021.3068206.

\bibitem{Elbir1}
A. M. Elbir and K. V. Mishra, "Deep Learning Design for Joint Antenna Selection and Hybrid Beamforming in Massive MIMO," 2019 IEEE International Symposium on Antennas and Propagation and USNC-URSI Radio Science Meeting, 2019, pp. 1585-1586

\bibitem{Sim} M. S. Sim, Y. Lim, S. H. Park, L. Dai and C. Chae, “Deep learning-based mmWave beam selection for 5G NR/6G with sub-6 GHz channel information: Algorithms and prototype validation,” IEEE Access, vol. 8, pp. 51634–51646, Mar. 2020.

\bibitem{Huang}H. Huang, J. Yang, H. Huang, Y. Song and G. Gui, “Deep learning for super-resolution channel estimation and DOA estimation based massive MIMO system,” IEEE Trans. Veh. Technol., vol. 67, no. 9, pp. 8549–8560, Sept. 2018.


\bibitem{Liu}
Z.-M. Liu, C. Zhang, and S. Y. Philip, “Direction-of-arrival estimation based on deep neural networks with robustness to array imperfections,” IEEE Transactions on Antennas and Propagation, vol. 66, no. 12, pp. 7315–7327, 2018.

\bibitem{quadriga}https://quadriga-channel-model.de/


\end{thebibliography}
\end{document}